\documentclass[12pt]{article}
\pdfoutput=1
\usepackage{amsmath,amssymb,amsthm,mathrsfs,bbm,caption,cite}
\usepackage{graphicx}
\usepackage[pdftex,colorlinks=true]{hyperref}
\usepackage[export]{adjustbox}
\usepackage[final,color]{showkeys} 
\definecolor{refkey}{gray}{0.75}
\definecolor{labelkey}{RGB}{155,48,48}
\renewcommand*\showkeyslabelformat[1]{%
  \fbox{\parbox[t]{0.8\marginparwidth}{\raggedright\normalfont\scriptsize\url{#1}}}}

\allowdisplaybreaks
\allowbreak

\usepackage{IEEEtrantools}
\usepackage[utf8]{inputenc}

\newcommand{\be}{\begin{equation}}
\newcommand{\ee}{\end{equation}}
\newcommand{\bea}{\begin{eqnarray}}
\newcommand{\eea}{\end{eqnarray}}
\numberwithin{equation}{section}
\newcommand{\ep}{\epsilon}

\newcommand{\nn}{\nonumber}

\newcommand{\mI}{\mathcal{I}}

\newcommand{\mO}{\ensuremath{\mathcal{O}}}
\newcommand{\OO}[1]{\ensuremath{\mathcal{O}\left(#1\right)}}

\newmuskip\pFqmuskip
\newcommand*\pFq[6][8]{%
  \begingroup 
  \pFqmuskip=#1mu\relax
  \mathcode`\,=\string"8000
  \begingroup\lccode`\~=`\,
  \lowercase{\endgroup\let~}\pFqcomma
  {}_{#2}F_{#3}{\left[\genfrac..{0pt}{}{#4}{#5};#6\right]}%
  \endgroup
}
\newcommand{\pFqcomma}{\mskip\pFqmuskip}

\newmuskip\Ftmuskip
\newcommand*\Ft[6][8]{%
  \begingroup 
  \pFqmuskip=#1mu\relax
  \mathcode`\,=\string"8000
  \begingroup\lccode`\~=`\,
  \lowercase{\endgroup\let~}\pFqcomma
  F_2{\left[\genfrac..{0pt}{}{#4}{#5};#6\right]}%
  \endgroup
}

\begin{document}
\begin{flushright}
\end{flushright}
~
\vskip5mm
\begin{center} 
{\Large \bf Eigenstate Thermalisation in the conformal Sachdev-Ye-Kitaev model: an analytic approach}
\vskip10mm

Pranjal Nayak\textsuperscript{a}, Julian Sonner\textsuperscript{b} \& Manuel Vielma\textsuperscript{b}\\
\vskip1em
\textsuperscript{a}Department of Physics \& Astronomy, University of Kentucky, 505 Rose St, Lexington, KY, USA\\
\textsuperscript{b}Department of Theoretical Physics, University of Geneva, 24 quai Ernest-Ansermet, 1211 Geneva 4, Switzerland
\vskip5mm

\textsuperscript{a}\tt{pranjal.nayak@uky.edu}\\
\textsuperscript{b}\tt{\{julian.sonner, manuel.vielma\}@unige.ch}

\end{center}

\vskip10mm

\begin{abstract}
The Sachdev-Ye-Kitaev (SYK) model provides an uncommon example of a chaotic theory that can be analysed analytically. In the deep infrared limit, the original model has an emergent conformal (reparametrisation) symmetry that is broken both spontaneously and explicitly. The explicit breaking of this symmetry comes about due to pseudo-Nambu-Goldstone modes that are not exact zero-modes of the model. In this paper, we study a version of the model which preserves the reparametrisation symmetry at all length scales. We study the heavy-light correlation functions of the operators in the conformal spectrum of the theory. The three point functions of such operators allow us to demonstrate that matrix elements of primaries ${\cal O}_n$ of the CFT$_1$ take the form postulated by the Eigenstate Thermalisation Hypothesis. We also discuss the implications of these results for the states in AdS$_2$ gravity dual.
\end{abstract}
\thispagestyle{empty}
\pagebreak
\pagestyle{plain}
\setcounter{page}{1}
\tableofcontents
\section{Introduction}
Understanding whether closed quantum systems thermalise, and if so in what precise sense \cite{DAlessio:2016rwt} occupies a central role; both in statistical physics, as well as holography, where aspects of thermalisation translate to detailed properties of black holes, their formation and evaporation, \cite{Horowitz:1999gf, Danielsson:1999fa, Chesler:2008hg, Bhattacharyya:2009uu, Anous:2016kss, Anous:2017tza}. In this work we analytically study a case that is of interest from both of these perspectives, the Sachdev-Ye-Kitaev (SYK) models, \cite{Sachdev:1992fk, Kitaev-talks:2015}. These are defined by a certain many-body quantum-mechanical Hamiltonian of $N$ complex or Majorana fermions interacting via a random $q$-body coupling, as we review in some detail below. This model, and the tensor-models that have a similar large $N$ behaviour, have an emergent reparametrisation symmetry in the infrared limit, \cite{Kitaev-talks:2015, Polchinski:2016xgd, Jevicki:2016bwu, Maldacena:2016hyu, Witten:2016iux, Gurau:2016lzk, Klebanov:2016xxf}. This emergent symmetry along with the chaotic behaviour of these models has motivated a study of a simpler version of AdS/CFT correspondence between a 1-dimensional quantum mechanical theory and a 2-dimensional theory of gravity that goes by the name nAdS$_2$/nCFT$_1$ correspondence, \cite{Kitaev-talks:2015, Jensen:2016pah, Engelsoy:2016xyb, Maldacena:2016upp, Cvetic:2016eiv, Mandal:2017thl}. There is by now a large corpus of work which studies this class of models, and which applies them: see \cite{Sarosi:2017ykf, Rosenhaus:2018dtp} (and the references therein) for recent reviews. A better understanding of the nAdS$_2$/nCFT$_1$ duality is important not only as a useful testing grounds to improve our understanding of the AdS/CFT correspondence and, by extension, quantum gravity in higher dimensions, it also describes the physics of near-extremal black holes in higher dimensions that have a near horizon AdS$_2$ geometry, \cite{Almheiri:2014cka, Nayak:2018qej, Larsen:2018iou, Castro:2018ffi, Moitra:2018jqs, Larsen:2018cts, Poojary:2018esz}. A major attraction of the SYK model (and its various cousins) is the ability to solve it analytically, especially in the infrared limit. This offers an opportunity to study the emergence of thermal behaviour in this system analytically. 

The eigenstate thermalisation hypothesis (ETH) provides a way to explain thermalisation in a quantum system from a microscopic point of view, \cite{Srednicki}. It states that with respect to a `typical' observable, individual eigenstates already contain all the information of the thermal ensemble. ETH can be encoded in terms of the matrix elements of a `typical' operator in the energy eigenbasis, 
\be\label{eq:ETHexpectation}
	\left\langle m \left| \mathcal O \right| n \right\rangle = \bar {\mathcal O}_{mc}\left(\bar E \right) \delta_{mn} + e^{-S(\bar E)/2} f(\bar E,\omega) R_{mn}
\ee
Here, $\bar{\mathcal O}_{mc}\left(\bar E\right)$ is the expectation value of the operator $\mathcal O$ in a microcanonical ensemble with average energy $\bar E$; $S(\bar E)$ is the corresponding microcanonical entropy of states with energy $\bar E$. Therefore, ETH requires that a typical operator be diagonal in the energy eigenbasis in the thermodynamic limit, while the off-diagonal terms are exponentially suppressed. In Random Matrix Theory (RMT), the canonical model of chaos and thermalisation in quantum mechanical systems, the off-diagonal matrix elements, $R_{mn}$, are randomly chosen from an ensemble with zero mean and unit variance.\footnote{In the notation of \eqref{eq:ETHexpectation} this means $f_{\rm RMT}(\bar E,\omega)=1$} However, for a more general theory, that is not necessary and could hold a key to understanding how the spectrum of different systems differs from that of RMT, \cite{Tolya}. 

This hypothesis has been verified for many quantum mechanical systems. However, the checks of this hypothesis in quantum systems have primarily been numerical in nature \cite{DAlessio:2016rwt, Garrison:2015lva}. Recent progress has been made in the analytic study of eigenstate thermalisation in holographic CFTs by \cite{Lashkari:2016vgj, Dymarsky:2016ntg, Lashkari:2017hwq, Basu:2017kzo, Faulkner:2017hll, Brehm:2018ipf}.
For a finite $N$ version of SYK model, eigenstate thermalisation was first established numerically in \cite{Sonner:2017hxc} by an exact diagonalisation of the Hamiltonian, and also in \cite{Haque:2017bts}. Eigenstate thermalization is also discussed in \cite{Lam:2018pvp} for Schwarzian quantum mechanics. In the current work, we analytically study eigenstate thermalisation in the IR limit of the SYK model. Some other related works that study thermalisation in SYK model are \cite{Kourkoulou:2017zaj, Eberlein:2017wah, Dhar:2018pii}.

In the IR limit, where the SYK model can be treated exactly and analytically, a reparametrisation symmetry emerges which is broken explicitly by the low-energy modes (pseudo-Nambu-Goldstones) of the model that are popularly known as \emph{Schwarzian modes}. The action for these modes is given in terms of the Schwarzian action,
\be
	\frac{\alpha}{J}\int d\tau \, {\rm Sch}\left(f(\tau);t  \right)~,
\ee
where, ${\rm Sch}\left(f(\tau);t  \right)$ denotes the Schwarzian derivative of the function $f(\tau)$,
\[
	{\rm Sch}\left(f(\tau);t  \right) = \frac{ f'''(\tau) }{ f'(\tau) } - \frac32 \left( \frac{ f''(\tau) }{ f'(\tau) } \right)^2~.
\]
It is this part of the full SYK action that breaks the reparametrisation symmetry explicitly. In this work, we study the correlation functions without the contribution of these low energy modes, whence the correlation functions are the conformal correlation functions. Through the operator-state correspondence a three-point function is related to an operator expectation value measured in energy eigenstates, see section \ref{sec:op-state}, and can be used to study eigenstate thermalisation.
\subsection*{Summary of results}
We find through a study of the three-point functions that the conformal sector of the SYK model thermalises via the mechanism of ETH. While what we describe here might look like an approximate computation in the SYK theory, a model where this is an exact computation of the correlation functions was recently discussed in \cite{Gross:2017vhb}. The corresponding two-dimensional dual theory is a quantum field theory of infinitely many massive scalar fields on the curved AdS$_2$ manifold. Since such a two-dimensional theory seems to lack gravitational modes, one might not expect to observe black hole formation. However, the observed thermalisation we find in our work leads to an interesting question of black hole formation in 2d gravity.

We also discuss the relation of the conformal sector of the SYK model to a generalised free theory (GFT) in one dimension and how our results on eigenstate thermalisation in the conformal sector of the SYK model also indicate the same in this GFT. This is a surprising result which we discuss in more detail in section \ref{sec:GFT}.


In a companion paper, \cite{Us-schwarzian}, we study detailed aspects of eigenstate thermalisation with the inclusion of the Schwarzian modes.
\subsection*{Plan of the paper}
We review the relevant properties of the SYK model in \autoref{sec:review}, summarising some key properties that we require in this paper while referring the reader to relevant papers for details. In \autoref{sec:GR}, we start with a discussion of spectrum of primary operators in the conformal limit of the SYK model. Then we describe the state-operator correspondence in a 1-dimensional theory and use it to define eigenstates created by insertion of primary operators. Thereby, we study the OPE coefficients in this theory, which are related to matrix elements of a light primary operator in basis of eigenstates. In the appropriate limit, these OPE coefficients describe eigenstate thermalisation in SYK. Later in \autoref{sec:GFT} we discuss eigenstate thermalisation in a generalised free theory in 1-dimensions. We conclude the paper with a summary and a plausible bulk interpretation of our results in \autoref{app:summary}. The appendices contain some supplementary details of our calculations.
\section{Relevant Properties of the SYK system}\label{sec:review}
In this section we summarise the main features of the Sachdev-Ye-Kitaev model relevant for our analysis. Much of this material is well known, but we wish to both make this work as self contained as possible and to draw the reader's attention to specific technical aspects we shall make use of later on. We keep the detail to the necessary minimum, but supply ample references which the interested reader is encouraged to follow up. We work mainly with the Majorana version of the model, to be introduced shortly and reviewed in more detail, for example in \cite{Sarosi:2017ykf, Kitaev:2017awl,Rosenhaus:2018dtp}. However, as will become clear, our results apply equally to the complex `Dirac' model \cite{Sachdev:2015efa}.  Without further ado, here is the Hamiltonian of the Majorana fermion SYK model generalised to $q$-body interactions,
\begin{equation}\label{eq.SYKHamiltonian}
H = i^{q/2} \sum_{i_1 < i_2 < \ldots < i_q \leq N} J_{i_1\ldots i_q} \psi_{i_1} \psi_{i_2}\cdots \psi_{i_q}\,.
\end{equation}
In this work we will focus our attention on the disorder-averaged theory, where the couplings $J_{i_1\ldots i_q}$ are averaged over a Gaussian random ensemble, with vanishing mean coupling and variance $\overline{J_{i_1\ldots i_q}^2}=J^2(q-1)!/N^{q-1}$. The disorder average gives rise to an $O(N)$ invariant theory, where the original $q$-Fermi interaction is squared to give an invariant $2q$-Fermi interaction with coupling strength $J^2$,
\begin{equation}
S = \int {\cal D}\psi{\cal D } G{\cal D} \Sigma \,\, e^{- S[\psi_i, G, \Sigma]}\,,
\end{equation}
with the averaged action $S$ written in terms of the original fermions, as well as two bilocal fields $G(\tau,\tau')$ and $\Sigma(\tau',\tau)$ whose role will become clear shortly,
\begin{eqnarray}
S &=&  \frac{1}{2}\sum_{i} \int d\tau\; \psi_i(\tau) ~\frac{\partial}{\partial \tau} \psi_i(\tau)  - \frac{J^2 N}{2 q} \int d\tau d\tau'  \;|G(\tau,\tau')|^q  \nn\\
 &&+ \frac{1}{2} \int d\tau d\tau' \;\Sigma(\tau',\tau) \left( N G(\tau,\tau') - \sum_{i}\psi_{i}(\tau) \psi_i(\tau') \right)\,,
\end{eqnarray}
The Lagrange multiplier field $\Sigma$ simply imposes the relation
\begin{equation}\label{eq.bilocalG}
G(\tau,\tau') = \frac{1}{N} \sum_{i=1}^N\psi_i(\tau) \psi_i(\tau')\,.
\end{equation}
The above action arises for the replica diagonal and the replica symmetric solution of the saddle point, \cite{1999PhRvB..59.5341P}. Alternatively, in the large $N$ limt, it can also be obtained by replacing the quenched disorder by an annealed disordered coupling, \cite{Kitaev-talks:2015, Maldacena:2016hyu}. In the subsequent discussion, we restrict ourselves to replica diagonal and symmetric ansatz which has the dominant contribution, \cite{Bagrets:2016cdf}.\\
This action is quadratic in the fermions, so that we can integrate them out to give rise to the usual ${\rm Tr}\log$ in the action. The result is an action written solely in terms of the bilocal {\it collective} fields $G$ and $\Sigma$, generating a vertex expansion that can be used to calculate the non-vanishing even-point functions of the fundamental fermions, via the relation \eqref{eq.bilocalG}. Note that both the collective fields are anti-symmetric under exchange of the time coordinates,
\be
	G(\tau,\tau') = -G(\tau',\tau), \qquad\Sigma(\tau,\tau') = -\Sigma(\tau',\tau)~.
\ee
\subsection{Vertex expansion}\label{sec.VertexExpansion}
 Applying this procedure we obtain the collective field action\footnote{We work in Euclidean signature.}
\begin{align}\label{eq.bilocalEffectiveAction}
-\frac{S_{\rm col}}{N} &= {\rm log}\,{\rm Pf} \left[ \delta(\tau-\tau')\partial_\tau + \Sigma(\tau',\tau)\right] + \frac{J^2}{2q} \int d\tau d\tau' \; |G(\tau,\tau')|^q \nn\\
&\hspace{7cm}-\frac{1}{2} \int d\tau d\tau'~ \Sigma(\tau',\tau)  G(\tau,\tau') \,.
\end{align}
The utility of this formulation is that it systematically determines integral (Schwinger-Dyson or `SD') equations for the $2n$-point functions of the fermions, order by order in a $1/N$ expansion.
To this end, let us note that this theory has a semi-classical limit as $N\rightarrow \infty$, governed by the saddle-point equations
\begin{subequations} \label{eq.SDExact}\begin{align}
&\delta(\tau-\tau')\partial_\tau + \Sigma(\tau',\tau) - G^{-1}(\tau,\tau') = 0 \\
&J^2 G^{q-1}(\tau,\tau')  - \Sigma(\tau',\tau) = 0
\end{align}\end{subequations}
\begin{figure}
\begin{center}
\includegraphics[trim={2cm 10cm 0cm 8.5cm},clip,width=0.9\textwidth]{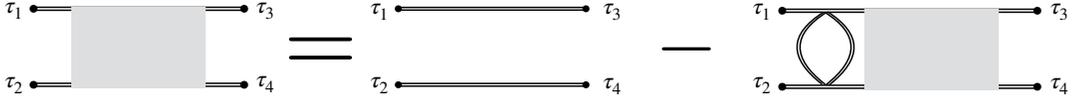}%
\caption{Schwinger-Dyson equation for the exact propagator ${\cal G}(\tau_1,\tau_2;\tau_3,\tau_4)$. For the theory formulated in terms of Majorana fermions this is a four-point function.\label{fig.FourPointSD}}
\end{center}
\end{figure}%
In order to generate the vertices relevant for the computation of fermion four- and six-point functions, we expand this action around the leading-order solution up to third order in the collective fields
\be\label{eq:S-col}
\frac{S_{\rm col}}{N} = S_{(0)} + \frac{1}{N}S_{(2)} + \frac{1}{N^{3/2}} S_{(3)} + \cdots\,,
\ee
where the term $S_{(2)}$ contains the two-point vertices in the collective fields and $S_{(3)}$ the three-point vertices, relevant for four- and six-point functions of the fundamental fermions respectively. More concretely, we follow a two-step procedure. We start by solving the first of the two SD equations to obtain
\be\
\Sigma(\tau',\tau) = -\delta(\tau-\tau')\partial_\tau + G^{-1}(\tau,\tau')
\ee
which we substitute back into the action to obtain
\be\label{eq.JeviciBilocalAction}
-\frac{S_{\rm col}}{N} = {\rm log}\,{\rm Pf} \left[G(\tau,\tau')\right] + \frac{1}{2}\int d\tau\left.\partial_\tau G(\tau,\tau')\right|_{\tau'\to\tau} + \frac{J^2}{2q} \int d\tau d\tau'~|G(\tau,\tau')|^q
\ee%
\begin{figure}
\begin{center}
\includegraphics[trim={3.5cm 8cm 2.5cm 6cm},clip,width=0.9\textwidth]{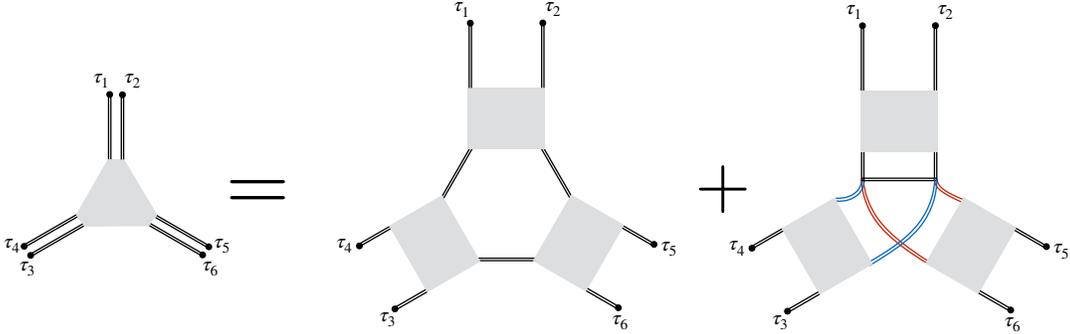} 
\caption{Schwinger-Dyson equation determining the three-point vertex ${\cal V}_{(3)}$. We note that there are two kinds of contributions on the right-hand side, planar ones and non-planar ones. For the theory formulated in terms of Majorana fermions this is a six-point function.\label{fig.SixPointSD}}
\end{center}
\end{figure}%
up to a constant which we have discarded. We then expand $G = G_{0} + \sqrt{\frac{2}{N}}\;g(\tau,\tau')$ to find the two and three-point vertices, starting with
\be \label{eq:S2}
S_{(2)} = \int d\tau_1 d\tau_2 d\tau_3 d\tau_4 \; g(\tau_1,\tau_2){\cal K}_{(2)}(\tau_1,\tau_2;\tau_3,\tau_4)g(\tau_3,\tau_4)\;,
\ee
allowing us to deduce the  propagator for the bilocal field
\be
{\cal G}(\tau_1,\tau_2;\tau_3,\tau_4) = \begin{gathered} \includegraphics[trim={2cm 10.35cm 19cm 8.5cm},clip,scale=0.4]{diagrams/FourPointSD.pdf} \end{gathered} \,.
\ee
and the Schwinger-Dyson equation shown in Figure \ref{fig.FourPointSD}. At the next order we find the three-point vertex
\be\label{eq:S3}
S_{(3)} = \int \prod_{i=1}^6 d\tau_i \; {\cal V}_{(3)}(\tau_1,\ldots \tau_6) \; g(\tau_1,\tau_2) g(\tau_3,\tau_4) g(\tau_5,\tau_6) \,,
\ee
where, once again the vertex succinctly encodes the Schwinger-Dyson equation shown in Figure \ref{fig.SixPointSD}, receiving both planar and non-planar contributions.
The precise mathematical expressions for these vertices are somewhat complicated and were originally computed in the papers \cite{Jevicki:2016bwu,Jevicki:2016ito,deMelloKoch:2018ivk}, and are reproduced in \autoref{app:col-act}. Note that the notation ${\cal K}_{(2)}$ and ${\cal V}_{(3)}$ is supposed to illustrate the fact that we should think of (the inverse of) ${\cal K}$ as a propagator for bilocal fields, while ${\cal V}_{(3)}$ gives a non-trivial three-point interaction vertex between bilocal collective fields.

\subsection{The IR theory \& Schwarzian contribution}
The discussion of the previous section is correct in the large $N$ limit, for all values of the coupling constant $J$. However the exact computation of the vertices requires the knowledge of the saddle point solution, $G_0$ (see \eqref{eq:K2} and \eqref{eq:V3}). The Schwinger-Dyson equations, \eqref{eq.SDExact}, are solvable in the deep infrared limit, $|\tau|J\gg1$, \cite{Kitaev-talks:2015, Polchinski:2016xgd, Jevicki:2016bwu, Maldacena:2016hyu}. In this limit, the derivative terms in the action \eqref{eq.bilocalEffectiveAction} and the SD equations \eqref{eq.SDExact} can be dropped. The resulting effective action \eqref{eq.bilocalEffectiveAction} has a reparametrisation (1D conformal) invariance under $\tau \rightarrow f(\tau)$ provided one transforms
\bea
G(\tau_1,\tau_2) &\rightarrow & f'(\tau_1)^{\Delta}f'(\tau_2)^{\Delta} G\left( f(\tau_1), f(\tau_2) \right)\,,\nn\\
 \Sigma(\tau_2,\tau_1) &\rightarrow &f'(\tau_1)^{1-\Delta}f'(\tau_2)^{1-\Delta} \Sigma\left( f(\tau_1), f(\tau_2) \right)\,.
\eea
Crucially this is not an invariance of the full action; working perturbatively around the IR limit, the leading action cost associated with a reparametrisation $f(\tau)$ can be   determined most efficiently by expanding \eqref{eq.JeviciBilocalAction} around the IR
\be
\frac{1}{2}\int d\tau\;\partial_\tau G(\tau,\tau')|_{\tau'\to\tau} = \frac{\alpha}{J}\int {\rm Sch}\left(f(\tau);\tau  \right)d\tau + \cdots\,,
\ee
where $\alpha$ is a coefficient that must be determined by solving the full Schwinger-Dyson equation \eqref{eq.SDExact}, which is typically done numerically, \cite{Kitaev-talks:2015, Maldacena:2016hyu}. Since this is the leading-order IR effect of breaking reparametrisations, we also refer to the Schwarzian contribution as the soft-mode action. In the current paper, we do not consider the effect of this soft-mode action on the correlation functions, which is discussed in our complementary paper, \cite{Us-schwarzian}. The exclusion of the Schwarzian modes leads to an exact conformal field theory. A modified version of the SYK model which has an exact conformal symmetry is discussed in \cite{Gross:2017vhb} and the results we present in this paper are applicable to this modified model as well. In the deep IR limit, where the SYK model has the approximate reparametrisation symmetry, its spectrum has a discrete tower of operators that are \emph{approximate} Conformal Primaries of $\mathbb {SL}(2,\mathbb R)$. We discuss the properties and correlation functions of the these operators in the next section. Our discussion follows the earlier work of \cite{Jevicki:2016bwu, Gross:2017hcz, Gross:2017aos}.
\section{Heavy States in Conformal Three-point Correlators}\label{sec:GR}
\subsection{Operators and states}
If we exclude the contribution of the Schwarzian mode we obtain an exact conformal field theory whose correlation functions obey the usual constraints imposed by $SL(2,\mathbb{R})$. Let us introduce the primary operators
\be\label{eq.op-spec}
{\cal O}_n (\tau_1) = \lim_{\tau_2\rightarrow\tau_1}\frac{1}{\sqrt N} \left[\sum_{k=0}^{2n+1} \sum_{i=1}^{N} d_{nk} \partial^k \psi_i (\tau_1)\partial^{2n+1-k}\psi_i(\tau_2)\right]\,,
\ee
which obey
\bea
\left\langle {\cal O}_m{\cal O}_n\right\rangle &=& \frac{\delta_{mn}}{|\tau_n - \tau_m|^{2h}}\,,\nonumber\\
 \left\langle {\cal O}_m{\cal O}_k{\cal O}_n\right\rangle &=& \frac{1}{\sqrt{N}}\frac{c_{mkn}}{|\tau_{mk}|^{h_m+h_k-h_n}|\tau_{mn}|^{h_m+k_n-h_k}|\tau_{nk}|^{h_n+h_k-h_m} }
\eea
with conformal dimensions
\be
h_n = 2n+1 + 2 \epsilon_n\, \ \ \epsilon_{n} = \frac{1}{q} \frac{2 n^2 +n + 1}{2 n^2 +n - 1}~, \  \ \ \ \ n\geq1~,\ \ \ q\gg 1~.
\ee
Using the 1D version of the operator-state correspondence, about which we  have more to say below, we can evaluate the matrix element of an operator ${\cal O}_k$ as a limit of the three-point function,
\be\label{eq.OperatorStateETH}
\langle E_m |{\cal O}_k | E_n\rangle=\left\langle {\cal O}_m(0){\cal O}(1)_k{\cal O}(\infty)_n\right\rangle = c_{mkn}\,.
\ee
This allows us to focus, as shown, on the OPE coefficient $c_{mkn}$. In the limit $m,n\gg k$, that is when the operators ${\cal O}_{m,n}$ create heavier states compared to the contribution of the probe operator ${\cal O}_k$,\footnote{see the discussion below for the precise notion of heaviness.} the OPE coefficients should satisfy the ETH relation, \cite{Lashkari:2016vgj},
\be\label{eq.ETHintermsofOPE}
c_{mkn} = f_k(\overline{E})\delta_{mn} + {\cal O}\left[e^{-S(\overline{E})/2}\right]\,,
\ee
where $f_k(\overline{E})$ is a smooth function of its argument $\overline{E} = \frac{E_m+E_n}{2}$ and the off-diagonal elements are suppressed by an entropic factor.
 Our objective is thus to compute the OPE coefficients using the bilocal action introduced above. This will evidently involve the computation of six-point functions of the fermions. Before embarking on the computation, we want to emphasise that the operators with $m,n\gg k$ are not \emph{heavy} operators in the conventional sense. Usually the term `heavy' is used to denote the operators whose dimension scales with the number of degrees of freedom of the theory ($N$ for vector theories and $N^2$ for matrix-like theories). This is consistent with the putative bulk dual description, where only fields massive enough to back-react on the geometry are expected to form black holes. However, the operators that we discuss in this work don't fall in this category. We consider operators whose conformal dimension approaches infinity but doesn't scale with $N$. We call these \emph{middleweight} operators. In higher-dimensional examples of the AdS/CFT correspondence such operators are dual to bulk fields that can be studied in the geodesic approximation, \cite{Balasubramanian:1999zv, Aharony:1999ti}, but would not by themselves, lead to a formation of a black hole. In the field theory description, that would mean that the states created by such operators don't thermalise. That we find any form of thermalisation for such operators is surprising. It is important to check this finding against a two-dimensional gravity analysis in future work.
 
 In the following section, we discuss our understanding of the state-operator correspondence in 1-dimension more precisely.

 \subsubsection{Comments on operator-state correspondence}\label{sec:op-state}
\begin{figure}
\begin{center}
\includegraphics[scale=0.6]{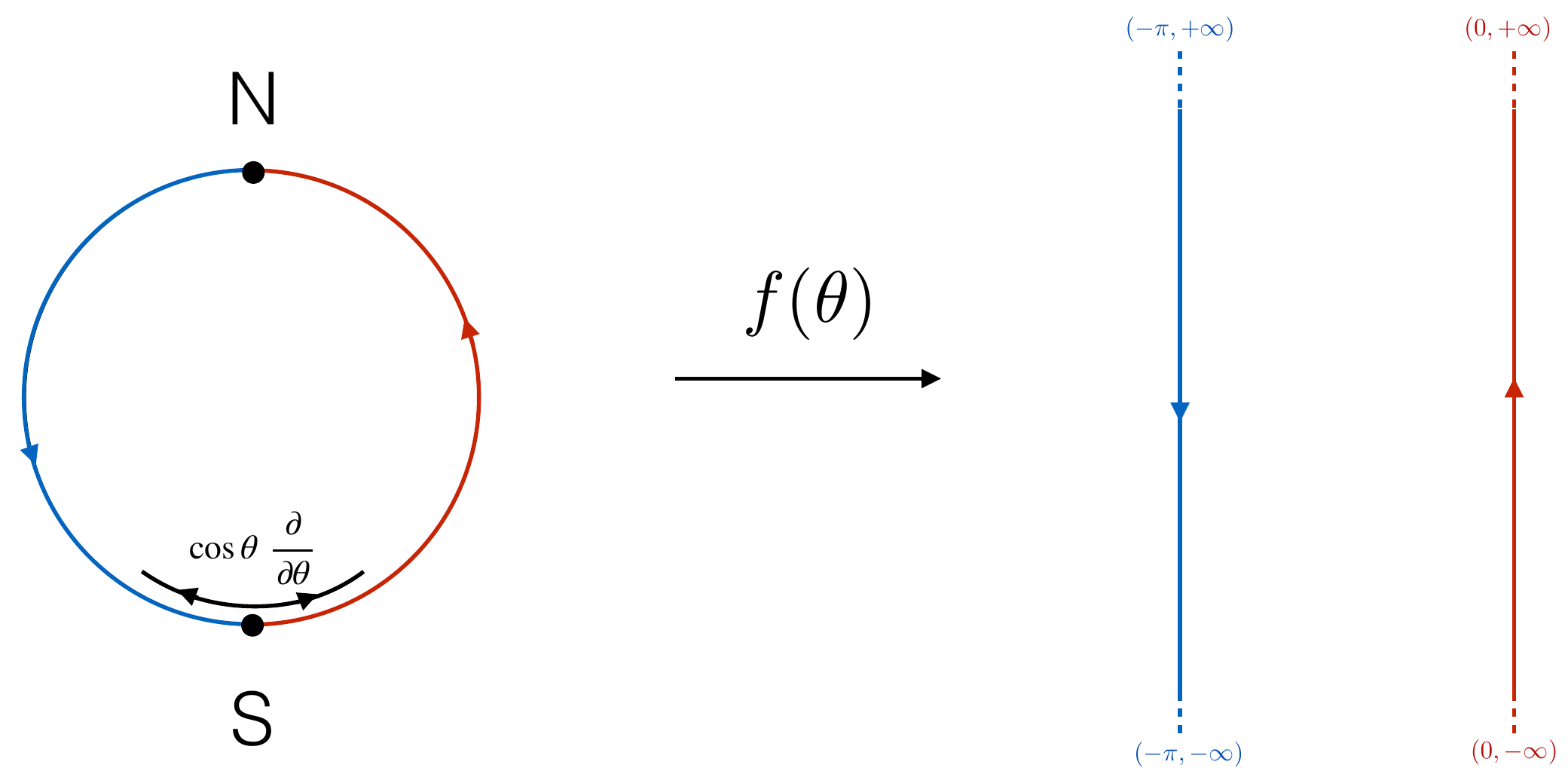} 
\caption{The operator-state map in 1D CFT proceeds via the map $S^1 \rightarrow S^0\times \mathbb{R}$ introduced in Eq. \eqref{eq.S1toS0crossRMap}. The semi circle in red is mapped to one copy of $\mathbb{R}$ shown in red, situated at $\sigma=0$, while the blue semi-circle is mapped to the blue copy of $\mathbb{R}$, situated at $\sigma = -\pi$. The Hamiltonian $\frac{\partial}{\partial\tau}$ is mapped to $\cos\theta \frac{\partial}{\partial\theta}$ and thus generates radial evolution as shown around the south pole. Modulo the subtlety of mapping to the disconnected $S^0\times \mathbb{R}$ this is the analog of radial quantisation in higher dimensional CFT.  \label{fig.OneDStateOperatorMap}}
\end{center}
\end{figure}

In dimensions $d\ge 2$ the operator state correspondence relies on the fact that we may conformally map the sphere $S^{d}$ to the cylinder $\mathbb{R}\times S^{d-1} $, so that asymptotic states at $\tau = \pm \infty$ on the cylinder map to operator insertions at the north and south poles of the sphere, or equivalently to the origin and the point at infinity of the plane $\mathbb{R}^d$. In one dimension this is more subtle, as the analogous map is one between $S^1$ and $\mathbb{R}\times S^0$, which corresponds to the geometry $\mathbb{R}\times \{0,-\pi \}$, i.e. two disjoint copies of the real line. This map is realised by setting \cite{Sen:2011cn}
\be\label{eq.S1toS0crossRMap}
\sigma + i\tau = 2 \arctan  \tanh \left(\frac{i\theta}{2} \right)
\ee
so that the two points $\theta=\pm \pi/2$ are mapped to $\tau = \pm \infty$. The segment $-\pi/2<\theta< \pi/2$ is mapped to a copy of $\mathbb{R}$ at $\sigma =0$, while the remaining half circle $\pi/2 < \theta < 3\pi/2$ is mapped to a second copy of $\mathbb{R}$ at $\sigma = -\pi$. The generator of time translations is mapped to $\partial_\tau = \cos\theta\partial_\theta$, which corresponds to 1D radial evolution as shown in Figure \ref{fig.OneDStateOperatorMap}.  In this way, a correlation function of the type \eqref{eq.OperatorStateETH} creates and probes a state in the doubled  Hilbert space ${\cal H}_{\rm CFT}\otimes {\cal H}_{\rm CFT}$. This has its manifestation on the bulk side where $AdS_2$ has a boundary consisting of two disconnected copies of the real line with time running in the same direction on both components.

\subsection{Fermion six-point functions and OPE coefficients}
In section \ref{sec.VertexExpansion} we explained how to generate the Schwinger-Dyson equation determining the six-point function of fermions from the bilocal action. We now proceed to actually evaluate the diagrams in the appropriate limit. As can be seen from Figure \ref{fig.SixPointSD}, there are two types of diagrams that contribute, {\it planar} diagrams and {\it non-planar} diagrams, also referred to as {\it contact} diagrams. Consequently, also the OPE coefficients can be organised into
\be
c_{nkm} = c^{(2)}_{nkm} + c^{(1)}_{nkm} \,,
\ee
where the first term corresponds to the sum of planar diagrams, and the second to the sum of non-planar diagrams. Consulting the definition of the ${\cal O}_n$ operators \eqref{eq.op-spec}, it becomes clear that the relevant limit of the fermion six-point functions is one where the insertions of the fermions approach each other in pairs. This allows one to use the OPE expansion of the fermions themselves,
 \be
 \frac{1}{N} \sum_{i=1}^N \psi_i(\tau_1) \psi_i(\tau_2) \mO_m(\tau) =  \frac{1}{\sqrt{N}} \sum_nc_n{\cal C}(\tau_{12},\partial_{\tau_2}) \mO_n(\tau_2)\mO_m(\tau) \,,
\ee 
with the result \cite{Gross:2017hcz}
\bea
	c^{(1)}_{mkn} &=& c_m c_n c_k \,  b^q (q-1)(q-2)\, \mI_{m k n}^{(1)}\label{eq:c1mkn}\\
	c^{(2)}_{mkn} &=& c_m c_n c_k\,  \xi_m \xi_n \xi_k\, \mI_{m k n}^{(2)}\,.\label{eq:c2nmk}
\eea
Here, $\mI_{m k n}^{(1)}, \mI_{m k n}^{(2)}$ are certain kinematic integrals arising in evaluating the contact and planar diagrams, respectively. The results of these integrals are given in \autoref{app:I1I2}.  In the large $q$ limit the $c_n$ are given by
\begin{equation}\label{eq.cn}
	c_n = \epsilon_n  \left[  \frac{n(2n+1)}{\left( \left(2n^2 + n + 1\right) \left(2n^2 + n - 1\right)   \right)}  \frac{\sqrt{\pi}  \Gamma(2n_1) }{2^{4n-2} \Gamma(2n+\frac{1}{2}) }  \right]^{\frac{1}{2}}\,,\qquad (q\gg 1)
\end{equation}
and the $\xi_n$ are defined as 
\be
\xi_n = \frac{2n+1}{2n^2+n+1}\,,\qquad (q\gg 1)\,.
\ee
In order to study the thermal properties of middleweight eigenstates,  we should take  a further limit, this time one where two of the operators, the ones creating the state, are much heavier than the remaining `probe' operator. We thus want to study the OPE coefficients $c_{nkm}$ in the limit $n,m\gg k$. This will allow us to simplify the corresponding expressions considerably. The first important simplification that occurs is the dominance of planer diagrams over contact diagrams $c^{(2)}_{nkm} \gg c^{(1)}_{nkm}$, as we show in detail in \autoref{app:behaviour_cijk}. We thus focus on $c^{(2)}_{mkn}$ in the remainder of this section. For our analysis we find it most convenient to work with the expressions for $c^{(2)}_{mkn}$ in the large $q$ limit, which was derived in \cite{Gross:2017hcz}, even though more explicit expression have been given later in \cite{Gross:2017aos}. The required kinematic integral can be written in the form
\be
{\cal I}_{m k n}^{(2)} = s_{m k n}^{(2)}     \left[2\frac{\ep_n^+ + \ep_m^-}{\ep_n^+ \ep_m^-}\frac{\ep_m^+ + \ep_k^-}{\ep_m^+ \ep_k^-}\frac{\ep_k^+ + \ep_n^-}{\ep_k^+ \ep_n^-} - \frac{1}{ \ep_m^- \ep_k^- \ep_n^- } - \frac{1}{ \ep_m^+ \ep_k^+ \ep_n^+}  \right]\,,
\ee
where $\epsilon_n^\pm = \epsilon_n \pm \Delta$.
Let us further define the variables
\be\label{eq:si-d-vars}
\sigma = n+m\,,\qquad d=n-m = \sigma \delta\,,
\ee
which express the average and difference of energies $\sigma$ and $\delta$ in the large $m,n$ limit.
Then the expressions for $s^{(2)}_{mkn}$ take the form of certain, combinatorial coefficients, which for arbitrary $k$ evaluate to
\be
s_{mkn}^{(2)} =\mathfrak{g}\left(4k-2 | 2k -2 \right) \frac{\Gamma \left( 2\sigma -1\right)}{\Gamma\left( \sigma - d \right)\Gamma(\sigma + d)}\,,
\ee
where $\mathfrak{g}\left(a|b\right)$ denotes a rational fraction (of variables $\sigma,d$) of degree $a$ in the numerator and degree $b$ in the denominator. We have tabulated the explicit form of these polynomials for the first few integer values of $k$ in \autoref{tab:snmk}.
\begin{table}[htp]
\footnotesize
\begin{center}
\begin{tabular}{|c|c|}
	\hline
	$k$ & $s_{mkn}^{(2)} $\\[5pt]
	\hline
	1 & $2 \left(d^2+\sigma^2+\sigma-1\right) \dfrac{ \Gamma (2 \sigma-1)}{\Gamma (\sigma-d) \Gamma (\sigma+d)}$\\[15pt]
	2 & $\dfrac1{3 (2s-2)(2s-3) } \bigg[7 d^6+d^4 \Big(9 \sigma^2 +9\sigma-53\Big)+d^2 \Big(9 \sigma (\sigma+1) \left(\sigma^2+\sigma-7\right)+118\Big)$\\
	& $\quad\qquad+(\sigma-1) (\sigma+2) \Big((\sigma^2+\sigma) (7 \sigma^2 + 7\sigma-36)+36\Big) \bigg]\times \dfrac{ \Gamma (2 \sigma-1)}{\Gamma (\sigma-d) \Gamma (\sigma+d)} $\\[15pt]
	3 & $\Bigg[33 d^{10}+45 d^8 \left(\sigma^2+\sigma-17\right)+d^6 \Big(10 (\sigma^2+\sigma) (5 \sigma^2 + 5 \sigma -103)+6459 \Big)$\\
	& $+5 d^4 \left((\sigma^2+\sigma) \left(10 \sigma^2 + 10 \sigma \left(\sigma^2+\sigma-22\right)+1681\right)-4887\right)$\\
	& $\qquad\qquad+d^2 \bigg(5 (\sigma^2+\sigma) \Big((\sigma^2+\sigma) \big((\sigma^2+\sigma) (9 \sigma^2+9\sigma-202)+1638\big)-5804\Big)+40308\bigg)$\\
	&$\qquad+3 (\sigma-2) (\sigma-1) (\sigma+2) (\sigma+3) \Big((\sigma^2+\sigma) \big((\sigma^2+\sigma) (11 \sigma^2 +11\sigma-157)+600\big)-600 \Big) \Bigg]$\\
	&$\dfrac1{30 (2\sigma-2)(2\sigma-3)(2\sigma-4)(2\sigma-5) }\times \dfrac{ \Gamma (2 \sigma-1)}{\Gamma (\sigma-d) \Gamma (\sigma+d)} $\\[10pt]
	$\vdots$ & $\vdots$\\
\hline
\end{tabular}
\end{center}
\caption{Table of $s_{mkn}^{(2)} $ evaluated for some small values of $k$}
\label{tab:snmk}
\end{table}%
We next notice that the ratio of Gamma functions is related to a regularised Kronecker delta symbol
\be\label{eq:reg-kronecker}
\mathfrak{b}_{\sigma,\delta} = 2^{-2(\sigma-1)}\sqrt{\pi\sigma} \frac{\Gamma \left( 2\sigma -1\right)}{\Gamma\left( \sigma - \sigma \delta \right)\Gamma(\sigma + \sigma \delta)} \quad\underrightarrow{\sigma\to\infty} \quad \delta_{\delta,0}
\ee
The width of the Binomial function, and consequently of the regularised Kronecker delta function\footnote{See \autoref{app:regular-delta} for a more in-depth discussion of the regularised Kronecker delta.}, in terms of $\delta$ is $1/\sqrt{\sigma}$, and approaches 0 as $\sigma \to \infty$. What this means in terms of the microcanonical ensemble is that we are considering all energy eigenstates within a window $\left(\sigma-\sqrt\sigma, \sigma+\sqrt\sigma\right)$. Such an interpretation of the microcanonical ensemble is conventional in the literature of Random Matrix theory, see \cite{DAlessio:2016rwt}. The limit, $\sigma\rightarrow\infty$ is the right limit to consider for our middleweight operators to be much heavier than the probe operator, and thus it is encouraging to see the Kronecker delta coming out. In fact, as we show now, the off-diagonal elements, corresponding to $n-m = d \neq 0$,  are suppressed by a factor $e^{-\sigma \ln 2}$. In order to extract the on- and off-diagonal functions in the ETH form of the matrix elements \eqref{eq.OperatorStateETH}, we may separate out the regularised Kronecker delta from the functions $c^{(2)}_{nkm}$, to obtain
\be
c^{(2)}_{mkn} = f_{mkn} \mathfrak{b}_{\sigma,d}.
\ee
We have now assembled all the ingredients that go into the evaluation of  $f_{mkn}$. Of course we are only interested in this quantity in the middleweight limit, which allows us to simplify  the full expression with the result
\be\label{eq.middleweightETHlimit}
\lim_{\sigma\rightarrow\infty}c^{(2)}_{mkn} = \frac{\Gamma(2k)}{2\sqrt{\Gamma(4k)}} \frac{k}{2k^2 + 2k +1} \left[  \frac{(2k+1) (2k^2 + k +1)}{(2k-1)(k+1)}   \right]^{3/2} p_k \sigma^{2k}\delta_{m,n} + e^{-\sigma\ln 2}R(\sigma,\delta)\,.
\ee
Here $p_k$ takes values $\left\{ 2,\frac{7}{12},\frac{33}{480}, \frac{715}{161280},\ldots  \right\}$, which are the leading coefficients in \autoref{tab:g}. This should be compared to the ansatz \eqref{eq.ETHintermsofOPE}. The result we have just found is precisely of the form appearing in the statement of eigenstate thermalisation, with a smooth function depending on the average energy on the diagonal and exponentially suppressed off-diagonal matrix elements depending smoothly on both the average energy as well as the difference. Isolating the diagonal contribution we find the expression
\be
\boxed{f_k(\overline E) = \frac{\Gamma(2k)}{2\sqrt{\Gamma(4k)}} \frac{k}{2k^2 + 2k +1} \left[  \frac{(2k+1) (2k^2 + k +1)}{(2k-1)(k+1)}   \right]^{3/2} p_k (2E)^{2k}}
\ee
where we have defined the average energy $\overline{E} = \frac{m+n}{2}$. We can also use Equation \eqref{eq.middleweightETHlimit} in order to obtain an explicit expression for the off-diagonal matrix elements
\begin{equation}
\boxed{R(\sigma,\delta) \!=\! \frac{1}{2}\left( 1-\delta^2  \right)^{\frac{1}{4}} \frac{\Gamma(2k)}{\sqrt{\Gamma(4k)}}\frac{k}{k^2 + 2k +1} \!\left[  \frac{(2k+1)(2k^2 + k + 1)}{(2k - 1)(k+1)} \right]^{\frac{3}{2}}\lim_{\sigma\rightarrow \infty}\mathfrak{g}(4k-2| 2k-2)}
\end{equation}
where we show the appropriate $\lim\limits_{\sigma\rightarrow \infty}\mathfrak{g}(4k-2| 2k-2)$ in \autoref{tab:g}.
\begin{table}[htp]
\begin{center}
\begin{tabular}{|c|c|}
	\hline
	$k$ & $\lim\limits_{\sigma\to\infty}\mathfrak{g}\left(4k-2 | 2k -2 \right)$\\[5pt]
	\hline
	1 & $2 \sigma^2 \left(1+\delta^2\right)$\\[5pt]
	2 & $\frac1{12} \sigma^4 \left( 7 \delta ^6+9 \delta ^4+9 \delta ^2+7 \right)$\\[5pt]
	3 & $ \frac{1}{480} \sigma^6 \left(33 \delta ^{10}+45 \delta ^8+50 \delta ^6+50 \delta ^4+45 \delta ^2+33\right) $\\[5pt]
	4 & $ \frac1{161280} \sigma^8 {\left(715 \delta ^{14}+1001 \delta ^{12}+1155 \delta ^{10}+1225 \delta ^8+1225 \delta ^6+1155 \delta ^4+1001 \delta ^2+715\right)} $\\[5pt]
	$\vdots$ & $\vdots$\\
\hline
\end{tabular}
\end{center}
\caption{Behaviour of $\mathfrak{g}\left(4k-2 | 2k -2 \right)$ in the $\sigma\to\infty$ limit}
\label{tab:g}
\end{table}%

Lastly, in the above discussion we showed that the off-diagonal matrix elements of the operator, ${\cal O}_k$, are suppressed with respect to the diagonal elements by a factor of $e^{-\sigma \ln 2}$. In the eigenstate thermalisation \emph{hypothesis}, the off-diagonal elements are suppressed exponentially by the microcanonical entropy at the average energy of the ensemble, $e^{-S(\bar E)/2}$. A naive counting of the number of states with energies between $\left(\bar E-\sqrt{\bar E},\bar E+\sqrt{\bar E}\right)$ gives $S(\bar E) \sim \sqrt{\bar E} + \ln(\bar E)$. Thus, we find that the observed suppression in the SYK model is stronger than what is required by ETH, together with our estimate $S(\bar E)$.  Our suppression although stronger than what is required according to the standard definition of ETH, does not violate the lower bound stemming from the requirement of ${\cal O}(1)$ fluctuations around the thermal value.\footnote{Requiring typical fluctuations of simple operators to be ${\cal O}(1)$, off-diagonal terms contribute a sum over $e^{S}$ terms, where $e^{S}$ is the dimension of the Hilbert space. Thus each term in the sum is at least $\mathcal O(e^{-S})$.} We leave an improved understanding of this mismatch for future work.

\subsubsection*{Summary}
In this section, we have shown that the states created by the conformal primaries in the IR limit of the SYK model, and more appropriately, in the cSYK model show Eigenstate thermalisation. This result may appear surprising from two different points of view. Firstly, the  dimensions of these operators do not scale with the degrees of freedom, $N$, in our theory (middleweight operators). From the bulk perspective these operators are dual to some fields that are not massive enough to back-react on the AdS$_2$, and yet the observed thermalisation would suggest that their insertion creates something like a blackhole microstate. We leave the exact bulk understanding of this observation to future work. It will also be  interesting to obtain the exact behaviour for truly heavy operators, i.e. ones that scale with $N$, whose insertion should change the leading saddle point in the bilocal collective field action \eqref{eq.JeviciBilocalAction}. Secondly, however, the result is surprising given that the OPE coefficients $c_{nkm}$ are dominated by the planar contribution, which can be shown to be closely related to those of a Generalised Free Theory. This point merits a  more in-depth discussion which we address in \autoref{sec:GFT}.

\subsection{Thermalisation in a generalised free theory?}\label{sec:GFT}
While it is satisfying to have explicitly established the thermality of eigenstates in the conformal sector of the SYK model at large $N$, our result is puzzling, seeing as it can be reproduced from a certain Generalised Free Theory (GFT). Following \cite{Gross:2017hcz} let us define the following GFT:
\begin{equation}
S = \frac{1}{2} \sum_{i=1}^N\int \frac{d\omega}{2\pi} \chi_i (-\omega) G_{(0)}(\omega)^{-1} \chi_i(\omega)\,,
\end{equation}
where $G_{(0)}(\omega)$ is the Fourier transform of the leading IR solution to the Schwinger-Dyson Equation \eqref{eq.SDExact},
\begin{equation}\label{eq.IRPropagatorFourier}
G_{(0)} (\omega) = i\cos\left( \pi\Delta  \right) \Gamma(1-2\Delta)|\omega|^{2\Delta - 1}{\rm sgn}(\omega)
\end{equation}
This theory is free, in the sense that all correlation functions can be obtained simply by Wick contractions, but it has a non-trivial propagator, given by \eqref{eq.IRPropagatorFourier}, and its primaries are given by the same expression \eqref{eq.op-spec}, now written in terms of the generalised free fermions $\chi_i$. One can thus obtain the three-point functions of ${\cal O}_n$ operators simply by repeatedly acting with derivatives as specified in the definition \eqref{eq.op-spec} and then taking the short-time limit. This results in the expression \cite{Gross:2017hcz}
\begin{align}
\left\langle {\cal O}_m {\cal O}_k  {\cal O}_n \right \rangle &=\underset{\tau_5\rightarrow \tau_6}{\text{lim}}\underset{\tau_3 \rightarrow \tau_4}{\text{lim}}\underset{\tau_1 \rightarrow \tau_2}{\text{lim}}\, \frac{1}{\sqrt{N}}\sum_{r=0}^{2n+1}\sum_{s=0}^{2m+1} \sum_{t=0}^{2k+1} d_{n r} d_{m s} d_{k t}\, \partial_1^r\, \partial_2^{2n+1-r}\nn\\
&\qquad \qquad \qquad \partial_3^s\, \partial_4^{2m+1-s}\, \partial_5^{t}\, \partial_6^{2k+1-t}
\Big[ G(\tau_1, \tau_6) G(\tau_2, \tau_3) G(\tau_4, \tau_5) + \text{perm}\Big] \nn \\
&= \frac{1}{\sqrt{N}}\frac{c^{free}_{mkn}}{|\tau_{mk}|^{h_m+h_k-h_n}|\tau_{mn}|^{h_m+k_n-h_k}|\tau_{nk}|^{h_n+h_k-h_m} }
\end{align}
from which we can extract the OPE coefficients themselves
\begin{align}
 c_{ m k n}^{free} &=  \frac{8\, s_{n m k}^{(2)}}{N_n^{free} N_m^{free} N_k^{free} }\\
&\text{where, } (N_n^{free})^2 =  \frac{2^{4n+1}}{(2n+1)  }\frac{\Gamma(2n+\frac{1}{2})}{\sqrt{\pi}\,  \Gamma(2n)}~. \nn
\end{align}
In the middleweight limit we are interested in, we can determine the ratio of the OPE coefficients in GFT and the SYK model,
\begin{align}
\lim_{m,n\gg k}\frac{c^{\rm SYK}_{mkn} }{c^{\rm free}_{mkn}} &= \lim_{m,n\gg k}
 \left[ 2\!\left(\! 1 + \frac{\epsilon_m^-}{\epsilon_n^+}\right)\! \! \left(1 + \frac{\epsilon_k^-}{\epsilon_m^+}\right)\!\! \left(1 + \frac{\epsilon_n^-}{\epsilon_k^+}\right)\! -\! 1 - \frac{\epsilon_n^- \epsilon_m^- \epsilon_k^-}{\epsilon_n^+ \epsilon_m^+\epsilon_k^+}\right] \nn \\
 & =  \frac{2 k^2+k+1}{2 k^2+k-1}~.
\end{align}
This tells us that our analysis, establishing that the SYK OPE coefficients take a form compatible with the eigenstate thermalisation hypothesis, also applies to the OPE coefficients of the GFT. On the face of it, this is very surprising; we usually do not think of free theories with correlation functions given in terms of Wick contractions as resulting in any kind of thermalisation. For the specific GFT under consideration here, however, this thermal behaviour results from the interplay of the non-trivial propagator and the combinatorics of the derivatives acting on the simple Wick-type six-point functions of the $\chi_i$ fermions. A similar analysis for a similar class of operators that were studied in \cite{Belin:2017nze} for a 2-dimensional CFT is discussed \autoref{app:Belin}. In that study of the OPE coefficients there we find lack of ETH like behaviour. The ETH behaviour for the generalised free theory described above is in some sense reminiscent of the thermal (chaotic) looking physics of the D1-D5 CFT at the orbifold point \cite{Balasubramanian:2016ids}, which is a free theory, albeit one with a complicated spectrum of operators. As a sanity check, one can quickly convince oneself that the limit in which the GFT actually  becomes the theory of free fermions ($\Delta = 0$), gives a vanishing three-point function, so we are saved from the conclusion that a theory of free fermions shows ETH itself.
\section{Summary and Discussion}\label{app:summary}
In this work, we have studied the 3-point correlation functions of the `excited' operators, given by \eqref{eq.op-spec}, that arise in the IR limit of the SYK model. In our analysis, we don't consider the contribution of the pseudo-Nambu-Goldstone (Schwarzian) modes to these correlation functions.   We argue that through a state-operator correspondence in a 1-dimensional theory, the OPE coefficients that appear in the three point function can be interpreted as matrix elements of an operator in the energy eigenbasis. By comparing these matrix elements in the thermodynamic limit with the expected behaviour \eqref{eq:ETHexpectation}, we argued that in this sector (without the contribution of the Schwarzian modes) the model shows eigenstate thermalisation. This is apparent from \eqref{eq.middleweightETHlimit}, where the expectation value of operator $\left\langle m\left|\mathcal O_k\right|n\right\rangle$ was studied in high-energy states, $m,n\gg k\sim \mathcal O(1)$ and it was argued that the operator becomes diagonal in this limit. We also computed the exponential suppression of the off-diagonal terms. While, our naive counting of microcanonical entropy at energy $\bar E$ gives $S\left(\bar E\right) \sim \sqrt{\bar E}+\ln\left(\bar E\right)$, we find a suppression by a factor of $\sim e^{-\bar E}$ which is stronger than $e^{-S(\bar E)}$. This suggests that our estimate of the microcanonical entropy might not be correct, or that the off-diagonal terms are indeed more strongly suppressed than is required for ETH. In any case, we do not expect the microcanonical entropy to grow faster than $\bar E$.

In the study of the Schwarzian modes to the correlation functions in an upcoming work, \cite{Us-schwarzian}, we find that the thermalisation in eigenstates depends upon the particular fashion in which the thermodynamic limit is taken. The analysis of that paper complements this paper's and presents some interesting observations of its own.
\subsection*{Future Directions}
Our current investigation has left us with some open questions that ask for a better understanding.
Above, we have discussed our lack of understanding of the density of middleweight states. This is important to understand the suppression of the off-diagonal terms compared to the diagonal terms. A better understanding of these terms will help us understand the departure of the model from the RMT behaviour, \cite{Tolya}.

We would like to study the correlation functions of the truly `heavy' operators (whose conformal dimension scales with $N$) in the conformal sector of the theory. Explicit computations involving such operators is often hard from the field theory point of view; and very little is known about such operators even in $\mathcal N=4$ super-Yang-Mills (sYM) theory. It might be possible that correlation functions of such operators are computable in the SYK model, even shedding some light on tools that could be used to do a similar computation for $\mathcal N=4$ sYM.

The current paper focused on the field theoretic study the SYK model in the conformal limit of the theory. The 2-dimensional dual, without the inclusion of the Schwarzian modes, is a theory of massive scalars on AdS$_2$ background. In future, we would like to understand this observed thermalisation from the bulk perspective in this non-gravitating theory of 2-dimensional gravity. A related question is to understand the chaotic behaviour of correlation functions in cSYK theory, both from the field theoretic as well as gravitational point of view. In our current understanding, the origin of chaotic behaviour in SYK model lies in the exchange of Schwarzian modes, \cite{Kitaev-talks:2015, Maldacena:2016hyu}. To our knowledge, no computation of out-of-time-ordered correlation functions has been performed in conformal sector of the SYK model (for the excited operators), and it will be illuminating to our understanding of thermalisation in cSYK model. 
\section*{Acknowledgements}
It is a pleasure to thank Dima Abanin, Alexandre Belin, Sumit Das, Anatoly Dymarsky, Nima Lashkari, Gautam Mandal, Shiraz Minwalla, Thomas Mertens, Charles Rabideau, Vladimir Rosenhaus, Ashoke Sen, Steve Shenker and Gideon Vos for insightful discussions and correspondence. JS would like to thank the organisers and participants of the Indian Strings Meeting 2019 as well as the workshop Qubits on the Horizon for feedback, where preliminary versions of these results were presented. PN acknowledges support from the College of Arts and Sciences of the University of Kentucky. PN also thanks the participants of PiTP Summer School at Princeton and Order from Chaos conference at KITP for the feedback on this work. MV would like to thank the ICTS Bengaluru and the TIFR Mumbai for their hospitality during the completion of this work. 
\appendix
\section{Collective field action}\label{app:col-act}
In this section we discuss the collective field action, \eqref{eq:S-col}. We derive the vertices of this action in an $1/N$ expansion around the solution of the replica diagonal and replica symmetric solution, $G_0(\tau_1,\tau_2)$, of the saddle point equations, \eqref{eq.SDExact}.\footnote{In addition to the replica diagonal saddle point solutions, we are also considering replica diagonal and symmetric fluctuations which are known to contribute dominantly, \cite{Bagrets:2016cdf}.} Using the expansion,
\be
	G(\tau_1,\tau_2) = G_0(\tau_1,\tau_2) + \sqrt{\frac2N} g(\tau_1,\tau_2)~,
\ee
in \eqref{eq.JeviciBilocalAction},
\be\label{eq.JeviciBilocalAction-app}
-\frac{S_{\rm col}}{N} = {\rm log}\,{\rm Pf} \left[G(\tau,\tau')\right] + \frac{1}{2}\int d\tau\left.\partial_\tau G(\tau,\tau')\right|_{\tau'\to\tau} + \frac{J^2}{2q} \int d\tau d\tau'~|G(\tau,\tau')|^q
\ee%
we get,
\bea
	S_{(2)} = \frac12 \int d\tau_1d\tau_2d\tau_3d\tau_4 \; g(\tau_1,\tau_2) \bigg[ G_0^{-1}(\tau_1,\tau_3)G_0^{-1}(\tau_2,\tau_4) \nn\\
	-3 J^2 G_0^{2}(\tau_1,\tau_2) \delta(\tau_3-\tau_1) \delta(\tau_2-\tau_4)\bigg] g(\tau_3,\tau_4)~.~
\eea
Comparing it with, \eqref{eq:S2}, we get,
\be\label{eq:K2}
	{\cal K}_{(2)}(\tau_1,\tau_2;\tau_3,\tau_4) = G_0^{-1}(\tau_1,\tau_3)G_0^{-1}(\tau_2,\tau_4) - 3 J^2 G_0^{2}(\tau_1,\tau_2) \delta(\tau_3-\tau_1) \delta(\tau_2-\tau_4)~.
\ee
Similarly, studying the fluctuations to the next order in $1/N$, one can compute the cubic interactions in the action,
\bea
	S_{(3)} = \sqrt2 \int d\tau_1\ldots d\tau_6 \; g(\tau_1,\tau_2) g(\tau_3,\tau_4) g(\tau_5,\tau_6) \bigg[ -\frac13 G_0^{-1}(\tau_6,\tau_1) G_0^{-1}(\tau_2,\tau_3) G_0^{-1}(\tau_4,\tau_5) \nn\\
	- J^2 G_0(\tau_1,\tau_2) \delta(\tau_3-\tau_1) \delta(\tau_4-\tau_2)\delta(\tau_5-\tau_1) \delta(\tau_6-\tau_2) \bigg]~.~
\eea
Once again, comparing with \eqref{eq:S3} gives us,
\begin{align}\label{eq:V3}
	{\cal V}_{(3)}(\tau_1,\ldots \tau_6) &= -\sqrt2 \bigg[ \frac13 G_0^{-1}(\tau_6,\tau_1) G_0^{-1}(\tau_2,\tau_3) G_0^{-1}(\tau_4,\tau_5) \nn\\
	&\qquad+ J^2 G_0(\tau_1,\tau_2) \delta(\tau_3-\tau_1) \delta(\tau_4-\tau_2)\delta(\tau_5-\tau_1) \delta(\tau_6-\tau_2) \bigg]~.~
\end{align}

\section{Dominance of Planar diagrams}\label{app:behaviour_cijk}
\begin{figure}[ht]
\begin{center}
\includegraphics[scale=0.52]{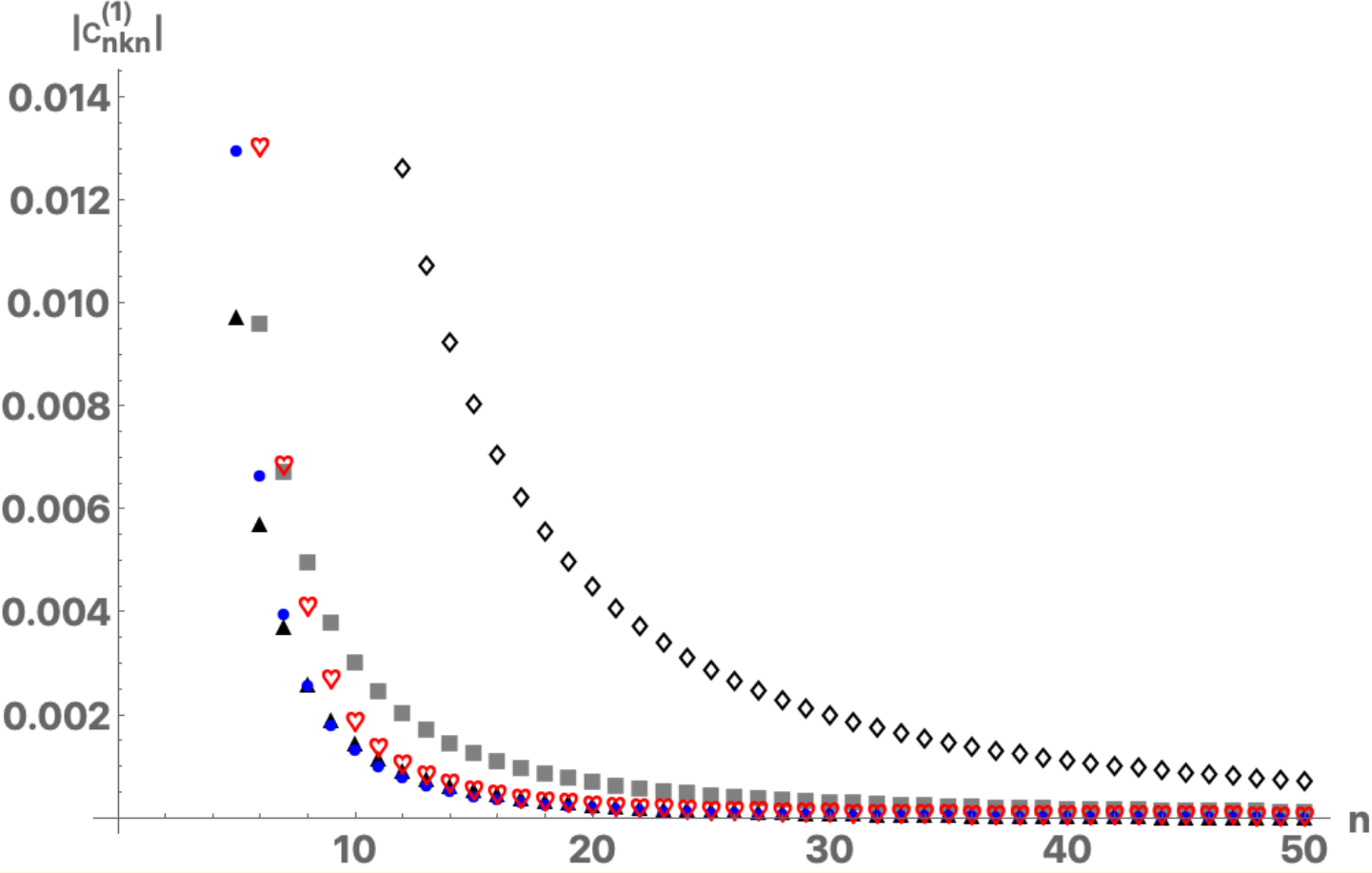} 
\includegraphics[scale=0.57]{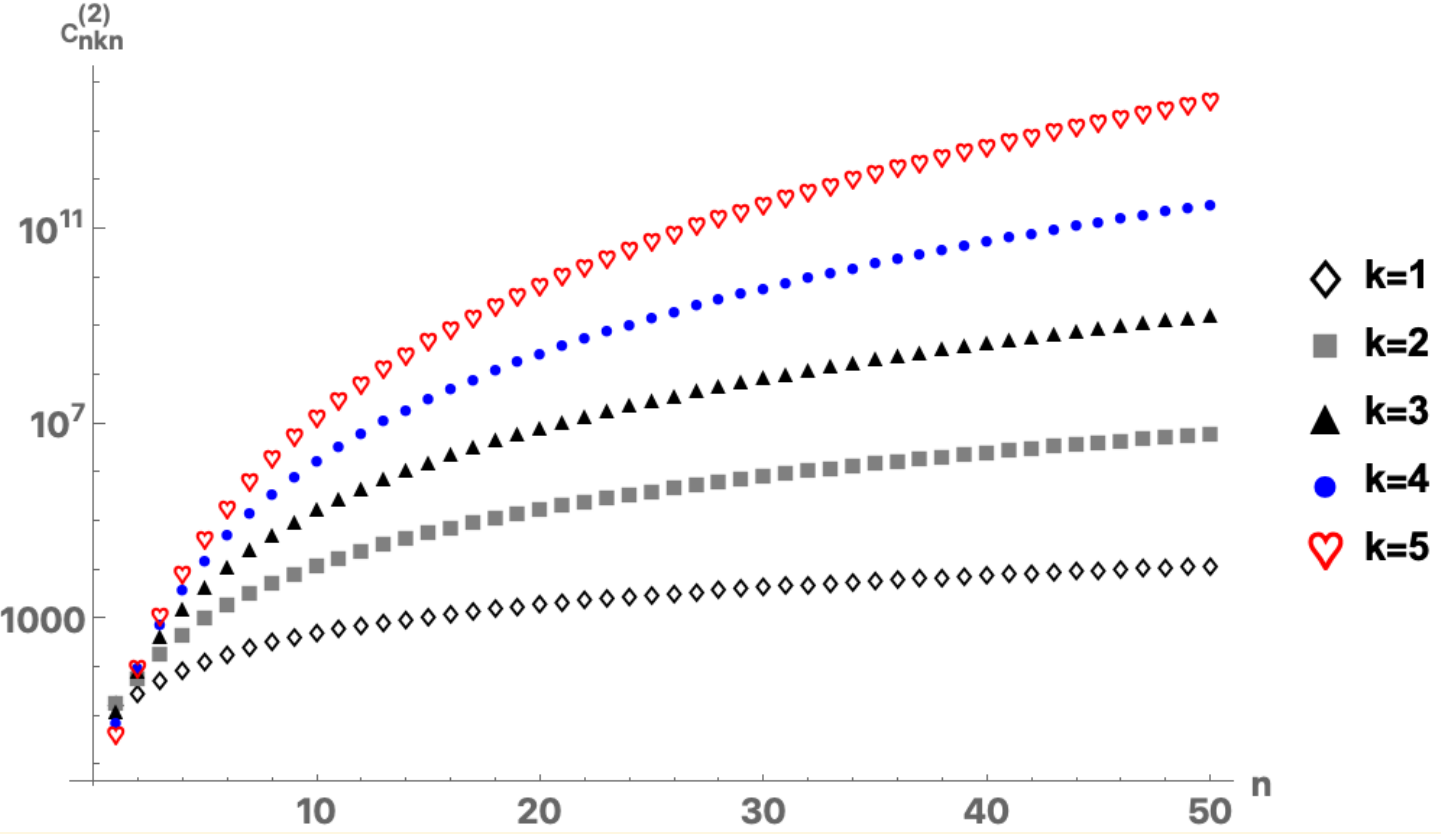} 
\caption{{\bf Left: }Diagonal values of the OPE coefficients $\left|c^{(1)}_{mkn}\right|$ become smaller as we take the limit $m,n\gg k$. {\bf Right: }Diagonal values of the OPE coefficients $c^{(2)}_{mkn}$ become larger as we take the limit $m,n\gg k$, thereby demonstrating the dominance of contact diagrams over the planar diagrams in this limit. Note that the y-axis of the right figure is plotted on a logarithmic scale. \label{fig.plotc1c2diag}}
\end{center}
\end{figure}

In this section, we demonstrate the dominance of planar diagrams over the contact diagrams in the limit in which two of the three insertions are much heavier than the third one. In \autoref{fig.plotc1c2diag} we plot the OPE coefficients, $c^{(1)}_{mkn}$ and $c^{(2)}_{mkn}$, (\eqref{eq:c1mkn}, \eqref{eq:c2nmk}) for $m=n$, and some small values of $k$ to demonstrate this dominance.

\begin{figure}[ht]
\begin{center}
\includegraphics[scale=0.52]{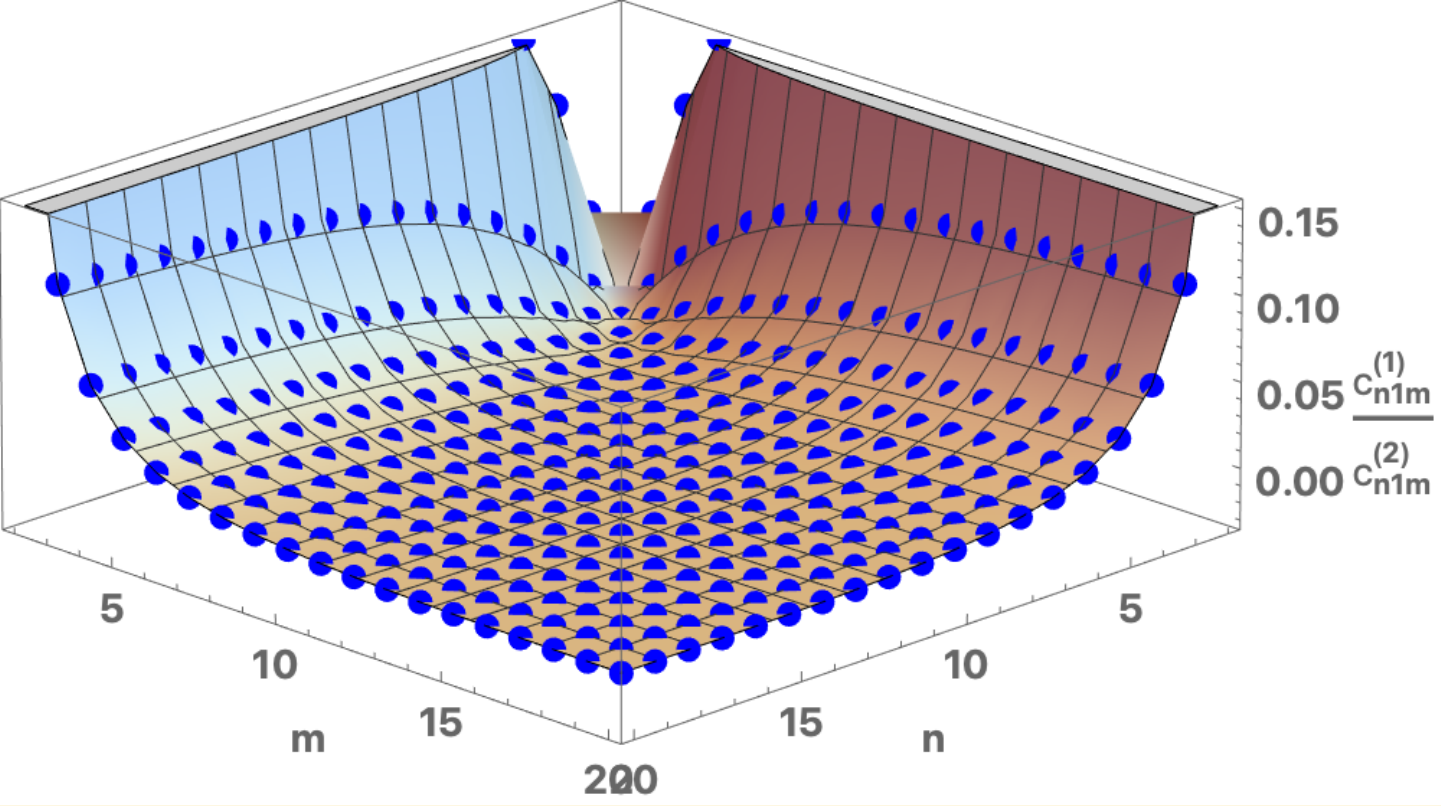} \hspace{10pt}
\includegraphics[scale=0.52]{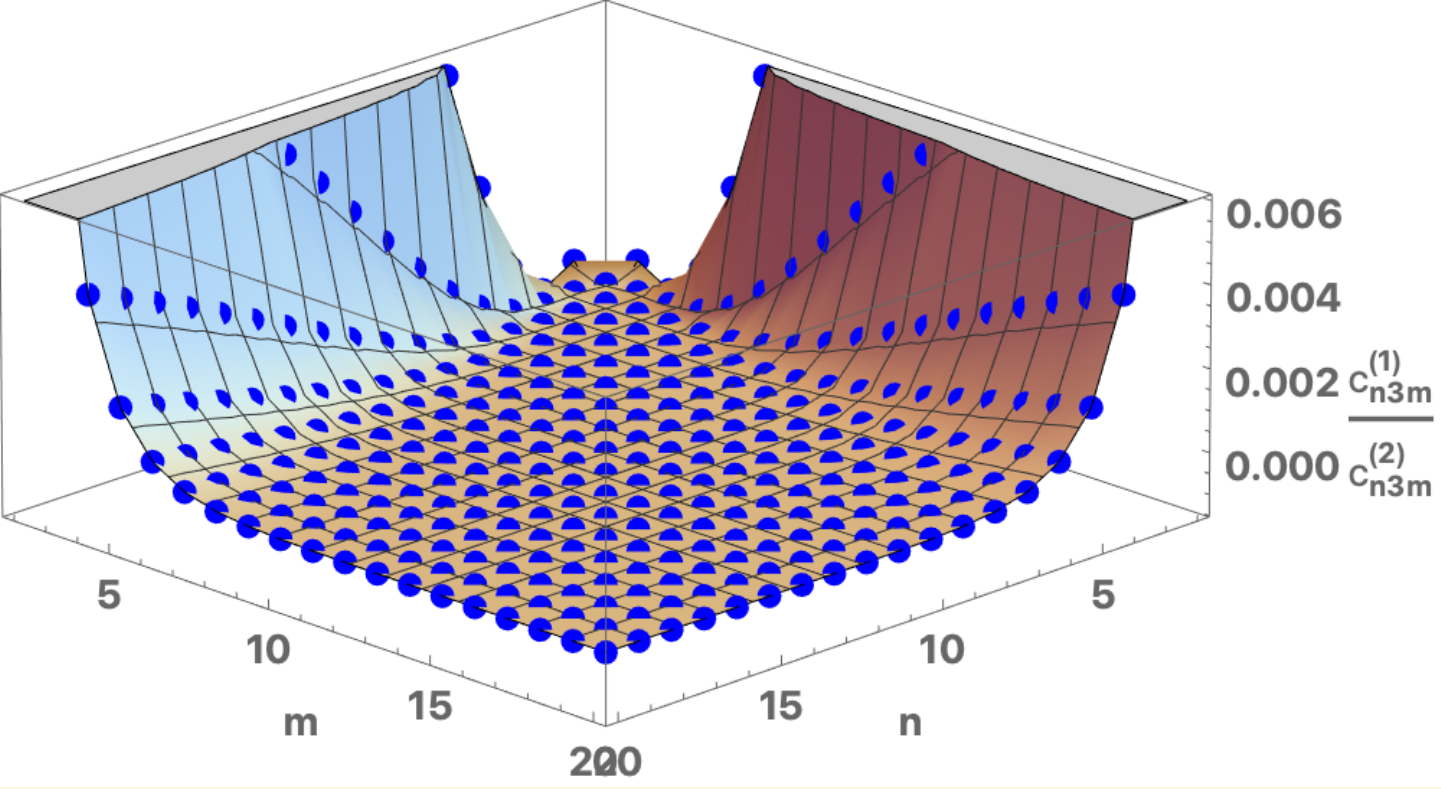} 
\caption{{\bf Left:} The ratio ${\left|c^{(1)}_{m1n}\right|}/{c^{(2)}_{m1n}}$ is largely suppressed even for $k=1$ as $m$ or $n$ are increased. {\bf Right:} Increasing $k$  accentuates the supression even further (here $k=3$).\label{fig.3Dk1plotc1byc2}}
\end{center}
\end{figure}

In \autoref{fig.3Dk1plotc1byc2}, we plot the ratio of the coefficients, ${\left|c^{(1)}_{m1n}\right|}/{c^{(2)}_{m1n}}$. This ratio is sufficiently small even for finite $m,n$. This plot also shows the suppression of the off-diagonal OPE coefficients corresponding to contact diagrams.

\section{Expressions for \texorpdfstring{$\mathcal{I}_{mkn}^{(1)}$}{I1} and \texorpdfstring{$\mathcal{I}_{mkn}^{(2)}$}{I2}}
\label{app:I1I2}
The exact expressions for $\mathcal{I}_{mkn}^{(1)}$ and $\mathcal{I}_{mkn}^{(2)}$ were derived in \cite{Gross:2017aos}. However, to avoid clutter and for the sake of intuition, we work with the corresponding expressions in the $q\gg1$ limit. In what follows, the notation $\mathcal{I}_{123}$ is used to refer to the term arising from a three-point function of operators of dimension $h_1, h_2$ and $h_3$, i.e. $\mathcal{I}_{123} \equiv \mathcal{I}_{h_1 h_2 h_3} \neq \mathcal{I}_{m=1, k=2, n=3}$.

\noindent For $q\gg 1$, the dimensions of the $\mathcal{O}_n$ approach their free-field values, $2n+1$, 
\begin{equation} \label{hnq}
h_n = 2 n + 1 + 2 \epsilon_{n}~, \ \ \ \  \epsilon_{n} = \frac{1}{q} \frac{2 n^2 +n + 1}{2 n^2 +n - 1}~, \  \ \ \ \ n\geq1~,\ \ \ q\gg 1~,
\end{equation}
while the OPE coefficients in the large $q$ limit behave as,
\begin{equation} \label{cnLarge}
c_n^2 = \epsilon_n^2 \frac{n(1+2n)}{\left(n(1+2n)+1\right)\left(n(1+2n)-1\right)}\frac{\sqrt{\pi} \Gamma(2n+1)}{\Gamma(2n+\frac{1}{2}) 2^{4n-2}}~, \ \ \ \ \ q \gg 1~.
\end{equation}

\noindent The contact diagram contribution to the three-point function $\langle \mathcal{O}_1 \mathcal{O}_2 \mathcal{O}_3\rangle$ has a coefficient that was denoted by $c_{123}^{(1)}$, given as  $c_{123}^{(1)} = c_1 c_2 c_3\, \mathcal{I}_{123}^{(1)}$ (see \eqref{eq:c1mkn}), where an exact expression for $\mathcal{I}_{123}^{(1)}$ can be found in \cite{Gross:2017aos}. Using (\ref{hnq}), an expression for $\mathcal{I}_{123}^{(1)}$ in the $q\to\infty$ limit is given by, \cite{Gross:2017hcz}, 
\begin{equation}\label{C1}
\mathcal{I}_{123 }^{(1)} = 2 s_{123 }^{(1)} \frac{\epsilon_1+ \epsilon_{n_2} + \epsilon_{n_3}}{\epsilon_{n_1} \epsilon_{n_2} \epsilon_{n_3}}~, \ \ \ \ q \gg 1~,
\end{equation} 
where $s_{123}^{(1)}$ is,
{\small
\begin{equation}\label{C1s}
\!\!s_{123 }^{(1)} =\!  (-4)^{n_1 + n_2 +n_3} \frac{\Gamma(\frac{1}{2}\!+n_2\!+\! n_3\! -\! n_1) \Gamma(\frac{1}{2}+n_1\! +\!n_3\!-\!n_2) \Gamma(\frac{1}{2}+ n_1\!+\!n_2\!-\!n_3) \Gamma(1\!+n_1\!+\!n_2\!+\!n_3)}{\pi^{\frac{3}{2}} \Gamma(1+2n_1 ) \Gamma(1+2n_2)\Gamma(1+2n_3)}  ~.
\end{equation} 
}%
The planar diagram contribution to the three-point function has a coefficient that was denoted by $c_{123}^{(2)}$, given in \eqref{eq:c2nmk} as $c_{123}^{(2)} = c_1 c_2 c_3\, \xi (h_1)\xi(h_2) \xi(h_3)\, \mI_{123}^{(2)}$. To avoid clutter, let us define, 
\begin{equation}
\epsilon^{\pm} = \epsilon \pm \Delta~,
\end{equation}
in terms of which, the factor $\xi (h_n)$ simplifies to, 
\begin{equation}
\xi(h_n)  =\frac{\epsilon^-}{\epsilon_n}\left( n + \frac{1}{2} \right)~,~\ \ \ \ \ \ q \gg 1~,
\end{equation}
while the expression for $\mathcal{I}_{123}^{(2)}$ is the $q\to\infty$ limit is,
  \begin{equation}  \label{I1232q}
 \mathcal{I}_{123}^{(2)} = s_{ 123}^{(2)} \left(2\frac{(\epsilon_1^+ + \epsilon_2^-)(\epsilon_2^+ + \epsilon_3^-)(\epsilon_3^+ + \epsilon_1^-)}{\epsilon_1^+\epsilon_1^-\epsilon_2^+\epsilon_2^-\epsilon_3^+\epsilon_3^-} - \frac{1}{\epsilon_1^+ \epsilon_2^+\epsilon_3^+}- \frac{1}{\epsilon_1^- \epsilon_2^-\epsilon_3^-}\right)~, \ \ \ \ q \gg 1~,
\end{equation} 
 with $\epsilon_{i} \equiv \epsilon_{n_i}$, and $s_{123}^{(2)}$ is,
  \begin{align}\label{s1232}
\!\!\! s_{123}^{(2)} &=  \frac{(2n_1+2n_2-2n_3)!(2n_2 +2n_3-1)!}{(2n_1-1)!(2n_2-1)!(2n_3-1)! (1+2 n_2 - 2n_3)!} \nn\\
&\qquad\qquad\qquad\qquad \times \pFq{4}{3}{1\!-\!2n_1~, 2\! +\! 2n_1~, 1\! -\! 2n_3~, -\! 2n_3}{ 2~, 1\!-\!2n_2\! -\! 2n_3~, 2\! +\! 2n_2 \!-\! 2n_3}{1}~,
\end{align}
\noindent where it is assumed that $n_1>n_2>n_3$. Using the definition of ${}_4 F_3$, this may be written as a single finite sum. Equation (\ref{I1232q}) was found by Gross and Rosenhaus in \cite{Gross:2017hcz} without taking the large $q$ limit of the exact answer, but rather by evaluating the integral $I_{123}^{(2)}$ to leading order in $1/q$. They noted that $s_{123}^{(2)}$ is the same expression that appears in computing the three-point function in a generalised free field theory with fermions of dimension $\Delta$, in the limit $\Delta\rightarrow 0$. Specifically,  
 \begin{align} \label{mCSum}
\!\!\!\!\! s_{123}^{(2)} = -\!\!\!\sum_{p_1, p_2, p_3}\!\! \!\!\binom{2n_1}{p_1}\!\binom{2n_2}{p_2}\!\binom{2 n_3}{p_3}\!
 \binom{2n_1\!+\!p_2\!-\!p_1}{p_2+1}\!\binom{2n_2\!+\!p_3\!-\!p_2}{p_3+1}\!\binom{2 n_3 \!+\!p_1\!-\!p_3}{p_1+1}\! \nn\\
\times \frac{z^{p_1-p_2 + 2n_2 - 2 n_3}}{ (\!-1\!-\!z)^{p_3 - p_2+2n_1 - 2 n_3}} ~,
 \end{align} 
where $z$ is a cross ratio of times; the answer is independent of $z$. While it is not manifest that (\ref{s1232}) and (\ref{mCSum}) are the same, one can verify that they are.

\section{Regularised Kronecker delta function}\label{app:regular-delta}
\begin{figure}[htbp]
\begin{center}
	\includegraphics[scale=0.5]{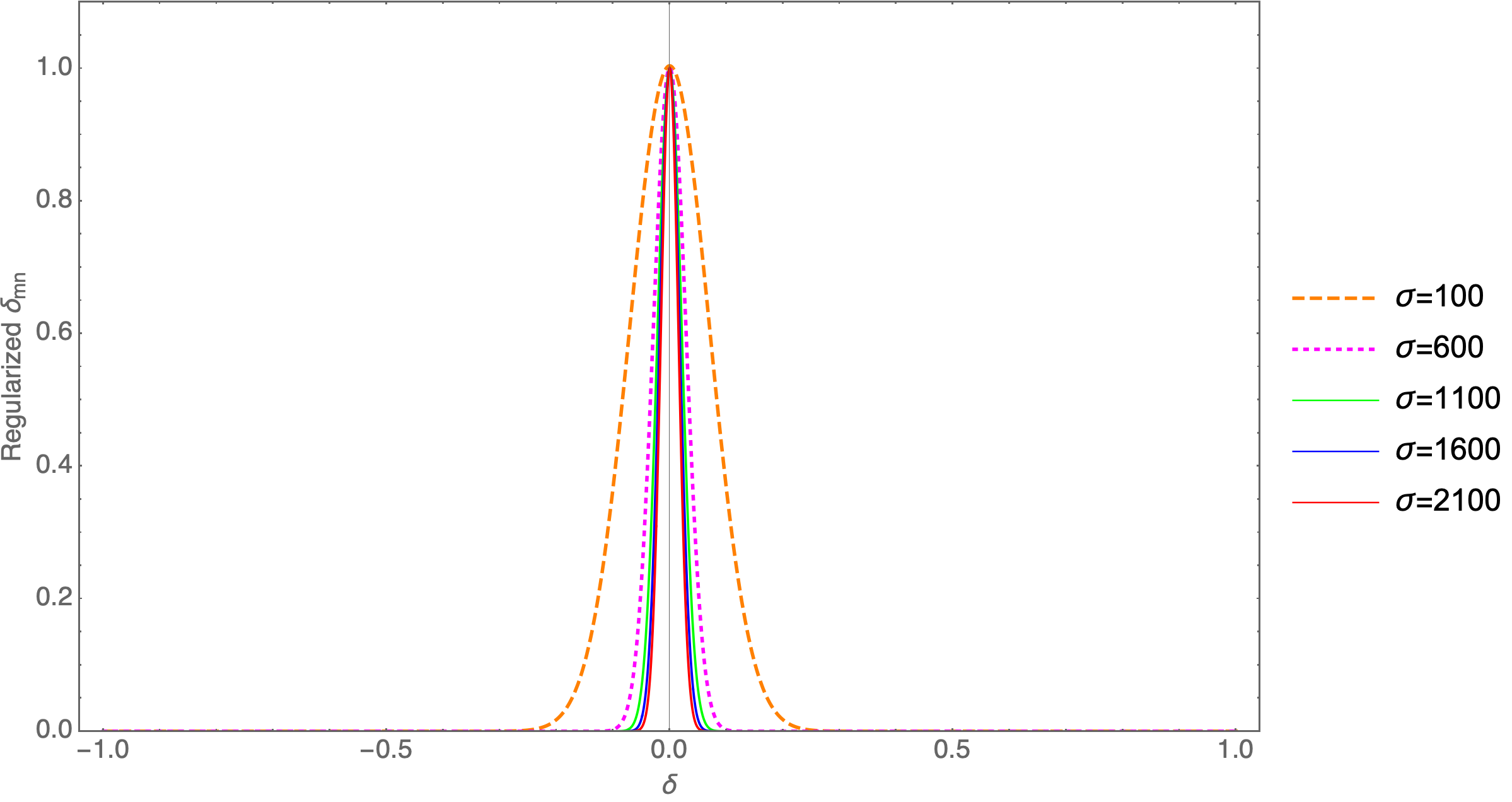} 
\caption{The function in \eqref{eq:reg-kronecker-app} evaluated for various values of $\sigma$.}
\label{fig:Kronecker-delta}
\end{center}
\end{figure}

Here, we discuss some of the properties of the regularised Kronecker delta function, in particular the exponential suppression of the off-diagonal components. Recall,
\begin{equation}\label{eq:reg-kronecker-app}
	\mathfrak b_{\sigma,\delta} = 2^{-2 (\sigma-1)} \sqrt{\pi  \sigma} \dfrac{ \Gamma (2 \sigma-1)}{\Gamma (\sigma-d) \Gamma (\sigma+d)} \quad \underrightarrow{\sigma\to\infty} \quad \delta_{d,0}~.
\end{equation}
\autoref{fig:Kronecker-delta} shows the behaviour of the function for increasing values of $\sigma$, after using \eqref{eq:si-d-vars}.
Notice, that the width of the plot decreases for increasing values of $\sigma$. The variable $\delta\in[-1,1]$ because the function on the LHS of \eqref{eq:reg-kronecker-app}, $\mathfrak b_{\sigma,\delta} $, isn't well defined outside this interval. However, we can define the function to be zero outside this interval. Let us next approximate the value of the function away from the peak. We do this by expanding $\mathfrak b_{\sigma,\delta} $ around, $\delta=1$:
\begin{equation}
	\mathfrak b_{\sigma,\delta} \sim 4^{-\sigma}\left[1 + (\delta -1) \sigma \left(\log \left(\frac{1-\delta }{2}\right)-1\right) + \OO{(1-\delta)^2}\right]
\end{equation}
Clearly the value of the function is suppressed exponentially for large values of $\sigma$.

\section{Lack of ETH in Free fields}\label{app:Belin}
In \cite{Belin:2017nze}, operators of the kind,
\begin{equation}
	O_k = :\partial^k T\partial^kT:~,
\end{equation}
were considered in a 2-dimensional conformal field theory. Here, $T$ is the stress tensor of the theory. The OPE coefficients for operators of this kind were computed there and was found to be,
\begin{equation}
	c_{k_1k_2k_3} = \sqrt8 ~\frac{ (k_1+k_2+3)!  (k_2+k_3+3)!  (k_3+k_1+3)!}{ (2k_1+3)! (2k_2+3)! (2k_3+3)! }~,
\end{equation}
where, $c_{k_1k_2k_3}$ is the OPE coefficient between the operators $O_{k_1}, O_{k_2}$ and $O_{k_3}$. To see if this shows ETH behaviour, \eqref{eq.ETHintermsofOPE}, let us factor out the regularised Kronecker $\delta$-function, \eqref{eq:reg-kronecker}. Then in the $k_1,k_3\to\infty$ limit, we get,
\begin{equation}
	c_{k_1k_2k_3} \approx \frac{\sqrt{\pi } 2^{-2 k_2-\frac{3}{2}} \left(4-\delta ^2\right)^{k_2} \sigma^{2 k_2+\frac{7}{2}}}{ \left(1-\delta ^2\right)^{\frac12} \Gamma (2 k_2+4)} \times \left( \frac{\left(1-\delta ^2\right) \left(\frac{1-\delta}{1+\delta}\right)^{-\delta } \left(\frac{2-\delta}{2+\delta}\right)^{\delta /2}}{4-\delta ^2} \right)^\sigma \mathfrak b_{\sigma,\delta}~.
\end{equation}
Here, once again we have used $k_1+k_3=2 \sigma$ and $k_3 - k_1 = d = \sigma \delta$. Note that unlike in \eqref{eq.middleweightETHlimit}, the function multiplying the regularised Kronecker $\delta$-function is exponential in $\sigma$. In fact, for diagonal elements corresponding to $\delta = 0$, this function is proportional to $e^{-2 \sigma\ln2}$. Consequently, as $\sigma \to \infty$, the diagonal elements vanish, thereby demonstrating lack of ETH-like behaviour.

\bibliographystyle{utphys}
\bibliography{refs3}

\end{document}